\documentclass[12pt]{iopart}
\usepackage{graphicx}

\begin{document}

\title{Time-Frequency Clustering for Burst Gravitational Waves Search in TAMA300 data}

\author{Ryota Honda, Shougo Yamagishi, Nobuyuki Kanda\\
 and TAMA collaboration}

\address{Graduate School of Science, Osaka City University,  Sumiyoshi-ku, Osaka 558-8585, Japan}

\ead{kanda@sci.osaka-cu.ac.jp}

\begin{abstract}
We have developed a method 'time-frequency (TF) clustering' to find the burst gravitational waves for TAMA data analysis.
TF clustering method on sonogram (spectrogram) shows some characteristics of short duration signal. 
Using parameters which represent the cluster shape, we can efficiently identify some predicted gravitational wave forms\cite{DFM}\cite{DFM2} and can exclude typical unstable spike like noises due to detector instruments.
The requirement of some parameters of cluster achieved roughly 50\% average efficiency for injected DFM waveforms of $h_{rss}\sim 2 \times 10^{-19}$ for type I burst. Also the reduction for signal by spike noises are more than one order improvement for the SNR$>$100.

\end{abstract}

\maketitle

\section{Introduction}

It is expected that stellar core collapses of supernova will radiate the burst gravitational waves,
which frequency is suitable to detect with ground-based gravitational wave detectors.
The predicted burst gravitational waves have a some peculiar characteristics : 
Type I) big narrow spike of $\sim$1 msec and following 10-15 msec decay ringing,
 Type II) a few msec spikes iterate a few times with $\sim$10 msec interval,
  Type III) broad excess of $\sim$5 msec with fine structure of $\sim$1 msec.
The predicted burst gravitational waves have a duration and tail of signals.
  Since the tail structure and frequency spread is not same as popular spike noises which 
 caused by the detector instruments, it will be possible to differentiate the gravitational
 waves from noises. However, it is hard to predict the gravitational wave forms in analytically
or hard to fit with a few parameters.

To identify such a signal, we have developed a new method by examining 'clustering' of signal powers in time-frequency domain\cite{hondaMthesis}. 
We study using TAMA observation data and numerical predicted gravitational waveforms by Dimmelmeier at al.\cite{DFM}\cite{DFM2}

In some other time-frequency domain analysis, which especially use 'cluster', were studied as reference \cite{Syl_TFC}, 
to find efficient power filter window. Our practical issue is employment much more parameters and find efficient
combinations of parameters.

\section{TAMA data and preprocess}

	 TAMA300 is a 300 m baseline laser interferometric gravitational wave detector 
in Mitaka, Tokyo, Japan\cite{TAMA}. The interferometer is Fabry-Perot-Michelson type with a power recycling technique. 
The strain sensitivity of the detector is $1.5 \times 10^{-21} {\rm [/\sqrt{Hz}]}$ around 1kHz, 
and the integral observation time is more than 3000 hours\cite{TAMA}\cite{TAMA_inspmgr}\cite{TAMAburst}.
The typical strain sensitivity of TAMA detector in 9th data taking (DT9) is displayed in figure 3 of the reference \cite{TAMAburst}. 
Gravitational waves from stellar-core collapse of supernovae, or quasi-normal mode oscillation of black holes,
has the typical time scale 10$\sim$100 msec. It corresponds to the good sensitive observation band 100Hz$\sim$1kHz of ground-based gravitational wave detectors as TAMA300.

\subsection{raw data and simulation data}

The time series signal from the TAMA interferometer is digitized by analogue-to-digital converter of 16 bit depth.
The sampling interval is 50$\mu$sec(=20kHz sampling). The raw data as time series signal $v(t)$ is stored in frame format\cite{FRAME}. One data file consists from 20 frames, and each frame is 3.2768 sec time series signal.
For the convenient of the event search analysis, we repack raw data frames into new frame of 52.4288 sec data for $2^{20}$ samples.
For this analysis, we use part of DT9 real data which has good stableness and better sensitivity. 

To investigate the search performance for burst gravitational wave, we used predicted burst gravitational waves by Dimmelmeier et al.\cite{DFM}\cite{DFM2} (DFM waveforms). We inject DFM waveform by software (off-line injection) on TAMA data.
 Since this signal $v(t)$ is given after the detector response, electric servo and whitening filter, we convert DFM waveform from
 strain $h(t)$ to $v(t)$ using transfer function, and embed to raw data. 

Also gaussian noise simulation data is used. Since real data include many
unstable spikes or glitch noises which have short time duration, these will make spurious signals of gravitational waves.
 Gaussian noise simulation data need to evaluate the statistical fluctuation of stable noises.

\subsection{Line removal}
Before transforming the time series data to time-frequency domain, we remove the line noises which come from
AC power line and its harmonics, mechanical resonances of suspension, etc. The detail is given in the references\cite{TAMAburst}\cite{TAMA_ALF}. 

\subsection{Short-FFT (sonogram)}

\begin{figure}[htb]
\begin{center}
\includegraphics[width=10cm]{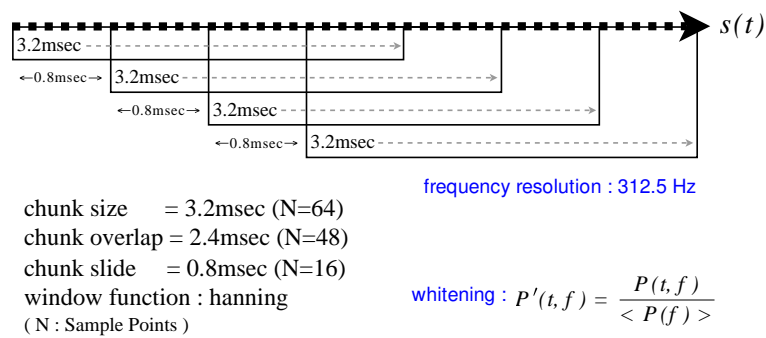}
\caption{\label{fig:TCHUNKS} Short-FFT (sonogram) process schematic}
\end{center}
\end{figure}

\begin{figure}[htb]
\begin{center}
\includegraphics[width=12cm]{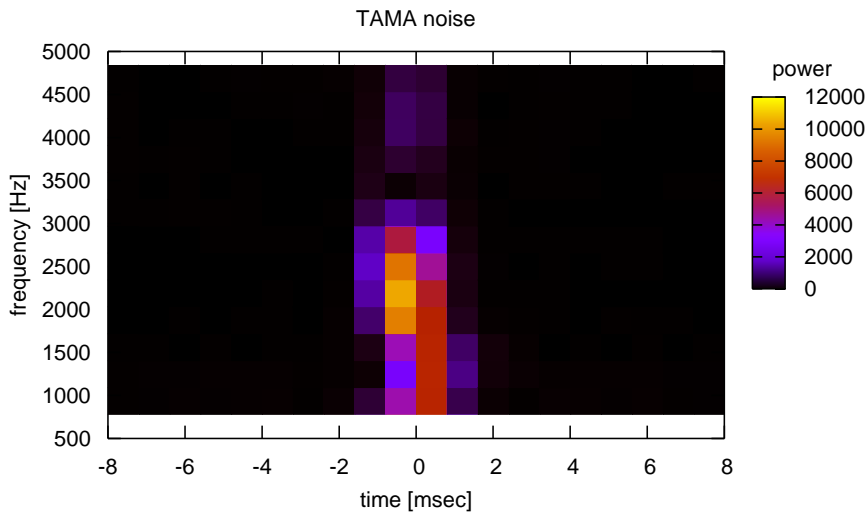}
\caption{\label{fig:SFFT1} sonogram example: TAMA spike like noise\\
Spike noises spread entire the frequency band, but concentrate in short time duration.}
\end{center}
\end{figure}

\begin{figure}[htb]
\begin{center}
\includegraphics[width=12cm]{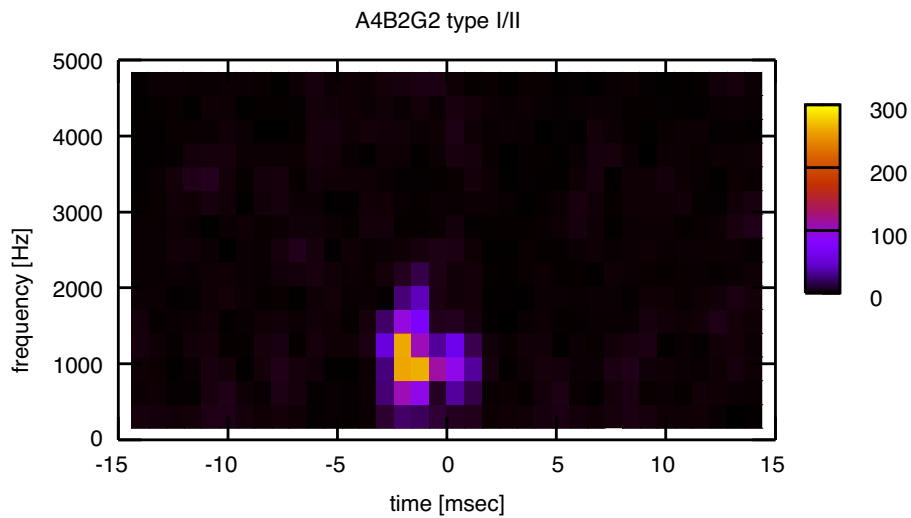}
\caption{\label{fig:SFFT2} sonogram example: burst gravitational wave (DFM waveform) injected TAMA data\\
Typical DFM waves not widely spread in frequency, but have a tail along the time axes.}
\end{center}
\end{figure}

We split time series data to 3.2 msec chunks. Time shift of next chunk is 0.8 msec, thus the
neighbor chunks overlap 2.4 msec.  The figure \ref{fig:TCHUNKS} displays this process.
For each of the chunks, signal 
  powers are calculated by using FFT (short-time Fourier transform,
  SFFT) with a Hanning window. The frequency resolution in the SFFTs 
  is 312.5Hz. 
  We dispose these SFFTs in time order, which is called 'sonogram' in time-frequency domain.
 One bin of data of time-frequency domain is displayed as 'pixel'.
Each 'pixel' represents the signal power at a given time and a frequency.
SFFT spectrum at each frequency band is normalized using average noise. 
Thus the sonogram will appear as flat pixels with gaussian noise.
 The figure \ref{fig:SFFT1} and  \ref{fig:SFFT2} are examples of sonogram of TAMA data around spike like noise, 
and typical DFM waveform injected in the TAMA300 data. Sonogram illustrate the pixel magnitude using color. 

\section{Clustering in time-frequency domain}

In time-frequency domain sonogram, we can find a closed pixels of which have higher magnitude in the figure \ref{fig:SFFT2} as 'cluster' of signal.

To identify and define a cluster are as follows:
1) We find all pixel which magnitude exceed signal-to-noise ratio $SNR > 3$. 
2) We assume that a local maximum pixel is a 'peak' of the cluster. The peak specified with its time $t_0$ and  frequency $f_0$. 
3) If one of a 8 pixels around the peak of the cluster exceed the cluster threshold, this pixel is 'connected' pixel. The cluster threshold is settled as square root of SNR of the peak pixel. 
4) We check the neighbor of connected pixels, again. 
5) If no more connected pixel cannot be found, we define the peak and these connected pixels as one 'TF-cluster'. 
6) We will iterate 2)-5) exclude the pixel which belongs to previous cluster, for other local peak.

\subsection{Characteristic parameters of clusters}
\begin{figure}[htb]
\begin{center}
\includegraphics[width=10cm]{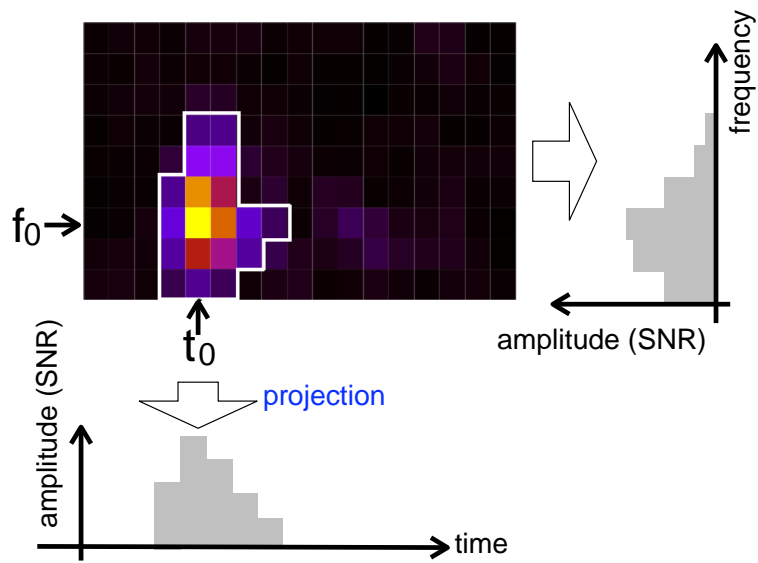}
\caption{\label{fig:cluster_profile} TF-cluster and its projection to each axes.\\
The pixels inside white border is recognized as one cluster.}
\end{center}
\end{figure}

When the cluster found, we calculate some parameters which shows cluster character.
Each cluster can be characterized by the absolute power, spread area, shape of power distribution et al.
Figure \ref{fig:cluster_profile} displays the schematic of projection of the cluster to each axes.
We can project the cluster by the numbers of pixels, or integrated magnitude of pixels along the orthogonal axes.
We employ following parameters to quantitate the shape of the projection; 
the number of pixels in the cluster for area $S$ ,
the sum of a magnitude of pixels for cluster volume $V$,
$tNs$ as $N$-th moment of projection of area (=the numbers of pixels) along the time axes,
$tNv$ as $N$-th moment of projection of magnitude along the time axes,
$fNs$ as same as a $tNs$ along the frequency axes,
$fNv$ as same as a $tNv$ along the frequency axes.

($t1s$, $f1s$) is a mean center of the cluster area, and  ($t1v$, $f1v$) is a center of magnitude of the 
cluster. Higher orders $tNs$ corresponds to the standard deviation, skewness et al.

\section{Event selection criteria}
\begin{figure}[htb]
\begin{center}
\includegraphics[width=10cm]{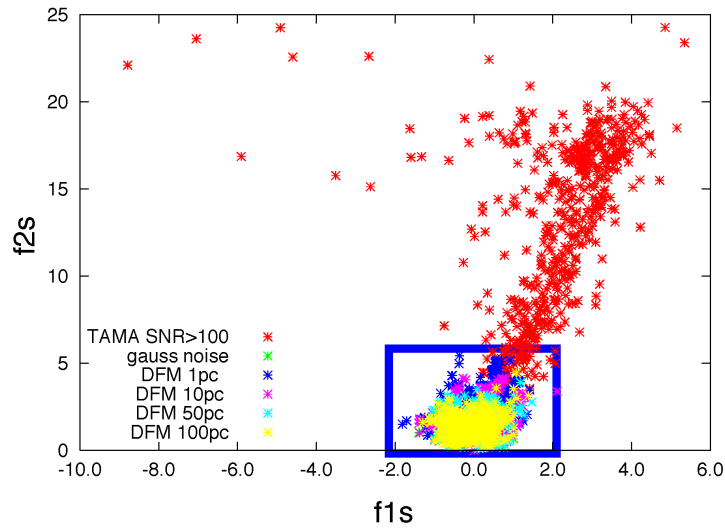}
\caption{\label{fig:paramscatt1} Scatter plot of cluster parameters $f1s$ - $f2s$\\
Red cross points is noises in real data. Others are injected gravitational waves (DFM waveforms) in simulation data.}
\end{center}
\end{figure}

\begin{figure}[htb]
\begin{center}
\includegraphics[width=10cm]{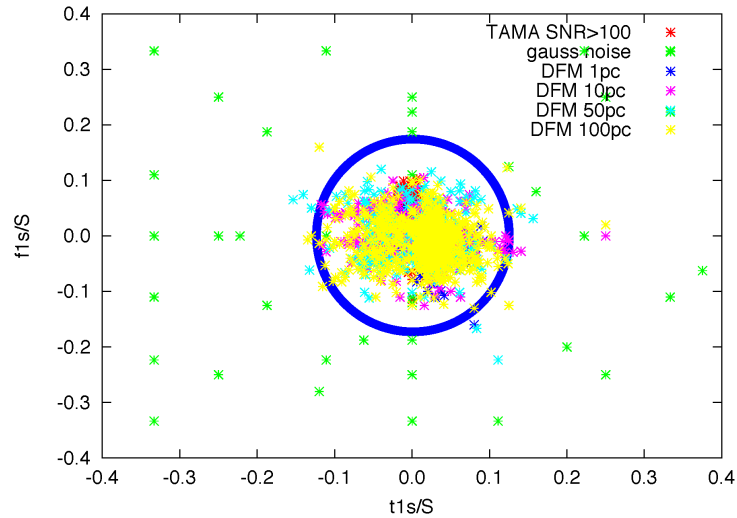}
\caption{\label{fig:paramscatt2}  Scatter plot of cluster parameters $t1s/S$ - $f1s/S$\\
Cluster due to gaussian noise fluctuation is spread uniformly in this figure.}
\end{center}
\end{figure}

Some combinations of the parameters makes possible to separate gravitational wave signal from many noises.
The figure \ref{fig:paramscatt1} and \ref{fig:paramscatt2} shows the example of two combination which efficiently separate the gaussian noise from the DFM wave forms.

In case of figure \ref{fig:paramscatt1} ($f1s$ - $f2s$ scatter plot), injected DFM wave forms concentrate within the box of blue solid line.
However, many TAMA noises spread wider region. Most of TAMA noises are spike noises.
One can see that the DFM waveforms can be efficiently separated from spike noises in the figure.
On the other, the figure \ref{fig:paramscatt2} shows the separation from gaussian noises.
Gaussian noises
will make spurious clusters accidentally according to the statistical fluctuations. Almost of them have small size, and parameters correlation is different from spike noise of gravitational waves.
In this $t1s/S$ - $f1s/S$ scatter plot, both spike noises and DFM waveforms concentrate in the blue circle. 

In this study, we require all the criteria in the table \ref{table:criteria} for the selection of burst gravitational wave candidates.

\begin{table}[htdp]
\caption{\label{table:criteria} Selection criteria of cluster parameters for burst gravitational waves}
\begin{center}
\begin{tabular}{c l c}
\br
parameters & brief meaning & criteria \\
& & (Unit is [pixels] except for $f_0$)\\
\mr
$S$ & area size of the cluster & $S \ge 4$\\
$f_0$ & frequency at the peak of the cluster & $f_0 \le 1250 \ {\rm [Hz]}$\\
\\
\multicolumn{3}{l}{\textit{For the frequency projection of the cluster;}}\\
$f1s$ & distance from center to the to the peak & $-2.0 \le f1s \le 2.0$\\
$f2s$ & standard deviation around $f1s$& $f2s \le 5.0$\\
$f4v$ & kurtosis around power weighted center $f1v$& $f4v \le 6.0$\\
\\
\multicolumn{3}{l}{\textit{For the time projection of the cluster;}}\\
$t1v$ &distance from weighted center to the to the peak & $-1.5 \le t1v \le 1.5$\\
$t2v$ &standard deviation around $t1v$& $t2v \le 3.0$\\
\\
\multicolumn{3}{l}{\textit{In combination;}}\\
$t2v,\ S$& &$\displaystyle \frac{\sqrt{t2v}}{S} \ge 0.04$\\
\\
$t1s^2,\ t2s^2,\ S$ & & $\displaystyle \frac{\sqrt{t1s^2 + t2s^2}}{S} \le 0.15$\\
\br
\end{tabular}
\end{center}
\label{default}
\end{table}%

\vspace{-0.5cm}

\subsection{Efficiency for burst gravitational waves}
With the criteria in the table \ref{table:criteria}, we estimate the efficiency by using Monte-Carlo simulation with random injection of  
the DFM gravitational waves in TAMA data. 
The figure {fig:distVSeff} shows the efficiency for TF-clusters of source distance from the earth. The efficiency depend on the type of burst waveform. Also the figure {fig:hrssVSeff} shows the efficiency for $h_{rss}$ : root square sum of strain amplitude.
The selection criteria for clusters which peak SNR$>$3 achieved 50\% average efficiency for injected DFM waveforms at 350 pc ($h_{rss}\sim 2 \times 10^{-19}$)
 for type I burst\cite{DFM},  and at 170 pc for all average.
 We take note that the efficiency looks higher than previous our analysis\cite{TAMAburst}, because we set lower SNR threshold
 for the comparison of clusters. If we wish to derive better upper limit of event search, we must set more higher threshold of SNR to reject small spurious noise.

\begin{figure}[htb]
\begin{center}
\includegraphics[width=10cm]{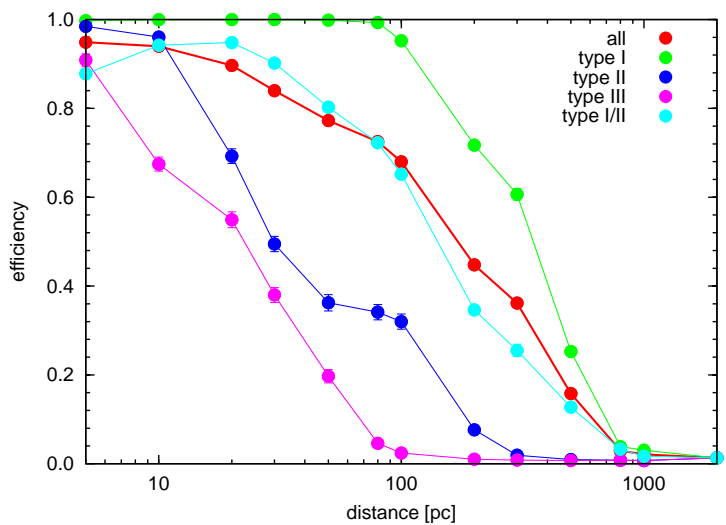}
\caption{\label{fig:distVSeff} TF-cluster selection efficiency for injected DFM waveforms (source distance)}
\end{center}
\end{figure}

\begin{figure}[htb]
\begin{center}
\includegraphics[width=10cm]{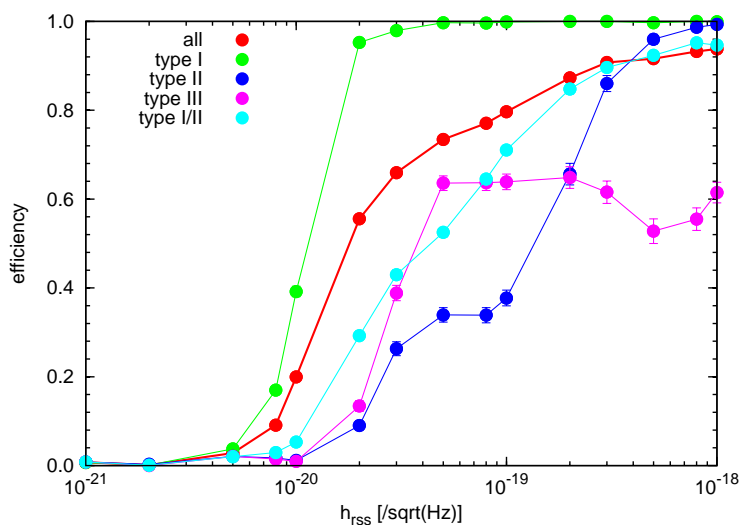}
\caption{\label{fig:hrssVSeff}  TF-cluster selection efficiency for injected DFM waveforms ($h_{rss}$)}
\end{center}
\end{figure}
\vspace{-0.5cm}

\subsection{Noise reduction}

TF-cluster selection criteria can reject typical noises efficiently in real data. Using a part of TAMA DT9 data
 ($1.26\times 10^5 sec$), we compare the SNR histogram of 'peak' before and after cluster parameter selection.
The reduction for spurious event clusters by spike noises are more than one order improvement for the SNR$>$100.
The reduction is efficient for higher SNR clusters for SNR$>$200. Because that TF-cluster method is efficient
 when the waveform is clear.

\begin{figure}[htb]
\begin{center}
\includegraphics[width=10cm]{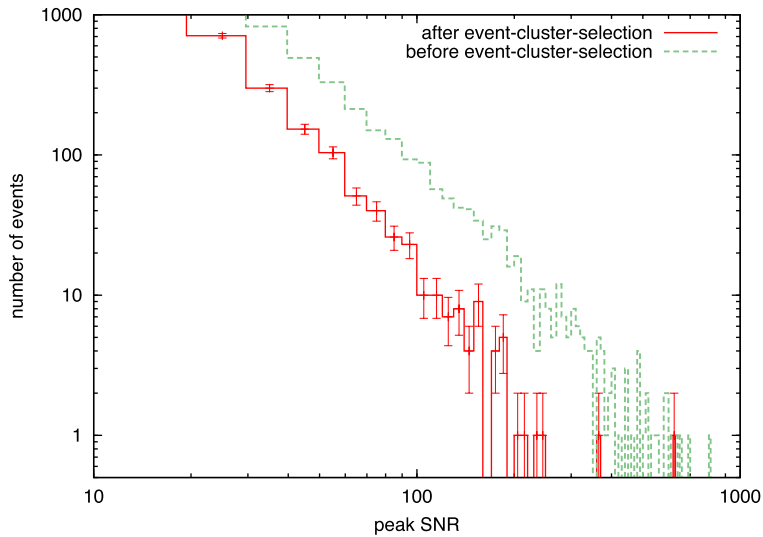}
\caption{\label{fig:trigHist} Histogram of SNR for real data\\
Histogram of cluster SNR is shown for a part of TAMA DT9 data. Green dashed line is all found cluster in TF domain. Red line is
after the selection by criteria in the table \ref{table:criteria}.}
\end{center}
\end{figure}
\vspace{-0.5cm}

\section{Summary and future plan}

We recently practice the newer search algorithm as TF-cluster using many cluster parameters. 
The method is efficient to select gravitational event and reject noise fake, especially in larger SNR case.
Since the TF-cluster method is not require the exact template of wave form, it is useful to identify the
burst gravitational waves which we believe only the typical characteristics is knows, or which predicted 
numerically without analytic waveform.

In the sonogram of this analysis, we employed SFFT, which has uniform time and frequency resolution. 
However, this is not optimal time-frequency resolution in an uncertainty principle of short wave packet.
According to the complemental relation between the time and the frequency, we can improve time resolution
 in high frequency band, can improve frequency resolution in lower frequency band instead of time resolution.
We began to employ with wavelet transform to achieve such resolution and to make more clear appearance of burst gravitational waves.

\section{Acknowledgments}
This research is supported in part by a Grant-in-Aid for Scientific Research on Priority Areas (415) of the Ministry of Education, Culture,Sports, Science and Technology.

\end{document}